\documentclass[11pt,twoside]{article}
\usepackage{atmp}
\copyrightnotice{2002}{6}{597}{617}	
\def\omit#1{}
\def\eql{~=~}

\def\coeff#1#2{\relax{\textstyle {#1 \over #2}}\displaystyle}
\def\half{{1 \over 2}}

\def\Tr#1{{\rm Tr} \left( #1 \right)}
\def\cA{{\cal A}} \def\cB{{\cal B}}
\def\cC{{\cal C}} \def\cD{{\cal D}}
 
 \def\cI{{\cal I}}
\def\cJ{{\cal J}} 
\def\cL{{\cal L}} \def\cM{{\cal M}}
\def\cN{{\cal N}} 
 
\def\cR{{\cal R}}

\def\bfone{\relax{\rm 1\kern-.35em 1}}

\def\us{\bf}

\def\IC{\mathbb{C}}

\def\IP{\mathbb{P}}

\def\IR{\mathbb{R}}
\def\nup#1({Nucl.\ Phys.\ $\us {B#1}$\ (}
\def\plt#1({Phys.\ Lett.\ $\us  {#1B}$\ (}
\def\plt#1({Phys.\ Lett.\ $\us  {#1B}$\ (}
\def\cmp#1({Comm.\ Math.\ Phys.\ $\us  {#1}$\ (}
\def\prp#1({Phys.\ Rep.\ $\us  {#1}$\ (}
\def\prl#1({Phys.\ Rev.\ Lett.\ $\us  {#1}$\ (}
\def\prv#1({Phys.\ Rev.\ $\us  {#1}$\ (}
\def\mpl#1({Mod.\ Phys.\ Let.\ $\us  {A#1}$\ (}
\def\ijmp#1({Int.\ J.\ Mod.\ Phys.\ $\us{A#1}$\ (}
\def\atmp#1({Adv.\ Theor.\ Math.\ Phys.\ $\bf {#1}$\ (}
\def\cqg#1({Class.\ Quant.\ Grav.\ $\bf {#1}$\ (}
\def\jag#1({Jour.\ Alg.\ Geom.\ $\us {#1}$\ (}
\def\jhep#1({JHEP $\bf {#1}$\ (}

\def\BZ{\mathbb{Z}}
\def\BR{\mathbb{R}}

\def\CL{\cL}

\def\CN{\cN}

\begin{document}
\title{Penrose Limits of RG Fixed Points and PP-Waves with Background
Fluxes} 
\url{hep-th/0205314}		
\author{Richard Corrado, Nick Halmagyi, Kristian D.\ Kennaway and
Nicholas P.\ Warner}		
\address{Department of Physics and Astronomy \\ 
and \\
CIT-USC Center for
Theoretical Physics\\
University of Southern California \\
Los Angeles, CA
90089-0484, USA}
\addressemail{rcorrado@usc.edu,halmagyi@usc.edu,kennaway@usc.edu,
warner@usc.edu}	
\markboth{\it PENROSE LIMITS OF RG FIXED POINTS\ldots}{\it R,\ CORRADO
ET AL.}

\begin{abstract}

We consider a family of pp-wave solutions of IIB supergravity.  This
family has a non-trivial, constant $5$-form flux, and non-trivial, (light-cone)
time-dependent RR and NS-NS $3$-form fluxes.  The solutions have
either 16 or 20 supersymmetries depending upon the time dependence.
One member of this family of solutions is the Penrose limit 
of the solution obtained by Pilch and Warner as the dual of a
Leigh-Strassler fixed point.  The family of solutions also provides
indirect evidence in support of  a recent conjecture concerning a large N 
duality group that acts on RG flows  of $\cN=2$ supersymmetric, quiver 
gauge theories.
\end{abstract}

\cutpage

\section{Introduction}

The remarkable observation in \cite{MetsaevBJ} that string theory is
solvable in 
certain pp-wave backgrounds has opened up new opportunities to test and analyze
string theory~\cite{BerensteinJQ}.  Combined with the observation that
pp-waves are Penrose   
limits of $AdS_m  \times S^n$ backgrounds, this has enabled extensive  new
testing to the AdS/CFT correspondence.  Our purpose in this paper is
to examine these issues for a supergravity solution that has already
provided a sharp set of tests of the AdS/CFT correspondence:  the
$\cN =2$ supersymmetric supergravity solution 
\cite{KhavaevFB,FreedmanGP,PilchEJ}  that  corresponds to a non-trivial, 
$\cN =1$ supersymmetric, conformal fixed point \cite{LeighEP} of $\cN =4$ 
supersymmetric Yang-Mills theory.  This supergravity solution 
involves non-trivial fluxes for the $3$-forms $H_{(3)}$, $F^{RR}_{(3)}$ and 
for the $5$-form, $F^{RR}_{(5)}$.  All of these fluxes remain non-zero in
the Penrose limit,  and indeed $H_{(3)}$ and $F^{RR}_{(3)}$ are 
non-constant.  The solution of \cite{KhavaevFB,FreedmanGP,PilchEJ}
 has ${1 \over 4}$-supersymmetry, and
we find that the corresponding pp-wave has 20 supersymmetries, which
may be understood as the ``universal'' 16 plus one quarter of the
supernumeraries.  We also find that, in spite of a ``light-cone
time-dependence,'' 
the string theory is still solvable in this background, and we compute the
modes.

The Penrose limit of the solution \cite{KhavaevFB,FreedmanGP,PilchEJ} 
is, in fact, one   point  in  a family of pp-wave solutions:   one can make 
more general  Ans\"atze for the fluxes by introducing arbitrary 
constant parameters, and one can even introduce a non-constant
dilaton and axion.  The result is a large, multi-parameter  space of solutions
that involves:  a constant,  symmetric, $8 \times 8$, real matrix in
the metric;  
a constant, complex, skew, $4 \times 4$ matrix in the $2$-form fields, 
$B^{NS} + i B^{RR}$; an arbitrary  complex constant, $a$, for the
dilaton/axion;   
an arbitrary real constant $f$ for $F^{RR}_{(5)}$;
and an arbitrary real parameter, $\beta$, that determines
the time dependence of the background fields.  The only non-trivial
equation is the $R_{++}$ Einstein equation, and it yields one real constraint.    
In this paper we will focus for simplicity on the family of solutions that have trivial 
dilaton and axion.  We will also specify a very particular form for
the $2$-form tensor gauge fields.  This form is motivated by the
Penrose limit of the solution \cite{KhavaevFB,FreedmanGP,PilchEJ},
but we will retain an arbitrary complex constant, $b$,  as the 
coefficient.  The Einstein equations then reduce to 
${1 \over 4} \beta^2 |b|^2 + f^2 = 1$,  and the solution space
is thus parametrized by $S^2 \times \IR$.
We find that these solutions have 20 supersymmetries if and only if
the time-dependence parameter is fixed to $\beta =- 2f$, and otherwise the
solution only has 16 supersymmetries.

The fact that the background $3$-form fields depend explicitly upon
$x^+$ means that the induced string action, in light-cone gauge, depends 
explicitly upon (world-sheet) time.  This time dependence naively
suggests   that  ``energy''  should not be conserved,  however, as one sometimes
finds in AdS backgrounds, it is natural to mix energy with angular momentum
to arrive at a more natural conserved ``energy.''    For the backgrounds that we 
consider here we find that the apparent time dependence of the
$3$-form fields can be removed by shifting an angular coordinate,
$\varphi \to \varphi + k\, x^+$, for some constant $k$,
to go to a co-rotating frame.  This
induces some off-diagonal metric terms, but the  metric remains
stationary and the background fields become independent of $x^+$.
For the Penrose limit of the background in \cite{PilchEJ} one has $k=3$
and the natural conserved energy is then:
\begin{equation}\label{twistedEn} 
\Delta ~-~ {3 \over 2}\, \cJ \,, 
\end{equation}
where $\Delta$ is the canonical AdS energy dual to the conformal
dimension of the operators on the brane, and $\cJ$ is the $R$-charge
of the $\cN=1$ field theory on the brane.  The combination,
\eqref{twistedEn}, is,  
of course,  the ``topological''  energy that vanishes on chiral, primary 
operators.

The family of theories described in this paper also provides some 
support for a conjecture made in \cite{CorradoWX}.  The goal of
\cite{CorradoWX}   
was to find the five-dimensional, gauged supergravity description 
of a class of holographic RG flows within $\cN=2$ quiver gauge theories.
For the $A_p$ quiver theory, this class of flows involves giving a mass
to the chiral multiplets associated to the nodes of the quiver, and so there are
 $(p+1)$ (complex) mass parameters.  The corresponding five-dimensional
gauged supergravity theory has an $SU(p+1)$ symmetry acting on  
the mass parameters, and so all of the flows are equivalent under this
symmetry.  At the non-trivial RG fixed point, an overall scale and phase
of these mass parameters disappears, leaving a fixed surface isomorphic
to $\IC\IP^{p}$.   The symmetry and the fixed-point manifold
 is manifest in the five-dimensional supergravity, but is very
surprising from the ten-dimensional perspective since it involves
trading topologically trivial $B$-field fluxes for K\"ahler moduli of
blow-ups of 
singularities.  In particular, there should be an $SU(2)$ symmetry acting
on an $S^2$'s worth of solutions that
smoothly interpolate between the flow of \cite{KlebanovHH} and the 
flow of \cite{FreedmanGP},   and even more simply,  there 
should be an  $SU(2)$ symmetry acting
on an $S^2$'s worth of solutions that smoothly interpolate 
between the IIB solutions corresponding to the RG fixed points,
that is between the $T^{(1,1)}$ solution of  \cite{RomansAN} and the 
solution of \cite{PilchEJ}.

While there have been some attempts to construct this $SU(2)$ family
of solutions directly, it is technically rather difficult \cite{CPW}.
Thus, a slightly more  
modest, but manageable,  test would be to see if the Penrose limits of
these solutions  
have the requisite $SU(2)$ symmetry.  The Penrose limit of the $T^{(1,1)}$
compactification was obtained and analyzed in
\cite{GomisKM, ZayasRX,ItzhakiKH},  and it was found
that the limit was the same as that of  the maximally supersymmetric
background  
of $AdS_5 \times S^5$, that is, the maximally supersymmetric
pp-wave of IIB supergravity \cite{MetsaevBJ,BlauNE}.  Thus the
Penrose limit seems to remove all of the supersymmetry-breaking
effects.   This does not happen with the Penrose limit of
the solution of \cite{PilchEJ}: as described above, the Penrose limit
has 20 supersymmetries, and is still in a sense, 
${1 \over 4}$-supersymmetric.   Thus, in the Penrose limit, the
$SU(2)$ symmetry and the $S^2$ conjectured in  \cite{CorradoWX} must 
smoothly interpolate between the Penrose limit of the solution of
\cite{PilchEJ} 
and the maximally supersymmetric pp-wave of IIB supergravity 
\cite{MetsaevBJ,BlauNE}.  At a generic point on the $S^2$ there must
be $20$ supersymmetries, and this will increase to $32$ when the background
$3$-form field vanishes.  This is, indeed, exactly what we find in
this paper.   
Moreover, it was also argued in \cite{CorradoWX} that the $SU(p+1)$ symmetry
should only be a large $N$ symmetry, and would be broken to
a discrete symmetry at finite $N$.  This means that it should be a 
symmetry of the supergravity, but not of the string spectrum.  Again,
this is confirmed by the results presented here.

Quite apart from the motivations of holographic field theory, the results
presented here are interesting in that they are backgrounds with
non-trivial, non-constant fluxes in which the supersymmetry
is {\it partially} broken, and yet the string theory is still solvable.
There have been quite a number of   pp-wave solutions in
which the string theory is solvable, and the supersymmetry
is broken using the metric, or by using  Penrose limits of intersecting branes 
(see, for example, \cite{BlauMW,GauntlettCS,CveticSI}).  In this paper the 
background $B$-field fluxes are intrinsically ``dielectric'': they are
not generated by a simple set of pure BPS branes.  When combined
with other families of solutions, the class presented here should
generate a new, richer class of interesting, supersymmetric pp-wave
solutions.

In section 2 we will introduce a general class of pp-wave
solutions to IIB supergravity, and then examine the Penrose limit of the
results of \cite{PilchEJ}.  In section 3 we will discuss the supersymmetry
of  the $S^2 \times \IR$ family of solutions described earlier.  Section
4 contains an analysis of the modes of the Green-Schwarz string in
an $S^2 \times \IR$ family of pp-waves, and section 5 contains some
final remarks.  Throughout this paper our conventions will be
those of \cite{SchwarzQR}, but with one minor exception: the complex
dilaton/axion field called $B$ in \cite{SchwarzQR} will be re-labeled 
$D_{(0)}$.

\centerline{\bf Note added in proof}

Since this paper was submitted to {\tt arXiv.org}, two closely related
papers~\cite{GimonSF,BrecherAR} appeared. Both of these papers
discuss aspects of the dual field theory, such as RG flows
and the operator spectrum, and they also take Penrose limits using other
geodesics.


\section{A class of pp-Wave solutions}

\subsection{The general class}

Consider a general pp-wave with an $\BR_r^4 \oplus \BR_y^4$ split, null
5-form,  complex 2-form potential, and a non-trivial dilaton and axion:
\begin{equation}\label{ppfluxsol}
\begin{split}
& ds^2 = 2\, dx^- \, dx^+   + \sum_{i<j} 
\left(A^{r}_{ij}\, r_i r_j + A^{y}_{ij}\, y_i y_j+ 2\, A^{m}_{ij}\,
r_i y_j \right) (dx^+)^2  \\
& \hskip1cm
- d\vec{r}^{\, 2} - d\vec{y}^{\, 2} \, , \\
& F=     f \, dx^+\wedge 
( dr_1\wedge\cdots \wedge dr_4 + dy_1\wedge\cdots \wedge dy_4 ), \\
& \cB = \coeff{1}{2} \, e^{i \beta x^+} \, \cC^{jk} \, 
dy_j \wedge dy_k \, ,\\ 
&D_{(0)}  =  a\, e^{2\, i \beta x^+}  \, . \\ 
\end{split}
\end{equation}
Following \cite{SchwarzQR}, we are using a metric that is ``mostly minus.''
The tensor, $\cC_{jk}= \cC^{jk}$, is a constant,  complex and skew symmetric
(we could, of course, extend $\cC$ into the other transverse directions,
but we are primarily interested in configurations that are Penrose
limits of IIB solutions  that have vanishing  $\cB $-components
in the $AdS_5$ directions.) The parameters $f$ and $\beta$ are
real constants, while $a$ is a complex constant.

Owing to the $x^-$--independence of the metric and the fact that
$g^{++}=0$, the only non-trivial IIB equation of motion \cite{SchwarzQR} is a 
component of the Einstein equation:
\begin{equation}\label{Rpp}
\begin{split}
R_{++}  ~&=~  \Tr{A^{r}+A^{y}}  \\  
&=~ 8 f^2 ~+~ 
(1-|a|^2)^{-1}\,\big(8\, \beta^2\,  |a|^2 \\
& \hskip 1cm ~+~  \coeff{1}{4} \, \big(\cC_{jk} - 
a\,\cC^*_{ jk} \big) \, \big(\cC^{*\,jk} - a^* \,\cC^{jk} \big) \big)
\, .
\end{split}
\end{equation}

For simplicity, we take a trivial dilaton/axion background, and
so set $a=0$.   For the string theory to be solvable in the background 
one needs to further restrict the matrix $\cC_{jk}$.  The choice that
we will make  
here\footnote{There are more general possible choices that lead 
to a solvable string spectrum.} is to take:
\begin{equation}
\label{specB}
  \cB = b \, e^{i \beta x^+} \, d\zeta_1 \wedge d\zeta_2\,,
\end{equation}
with $\zeta_1 = y_1 + i y_2, \zeta_2 = y_3 + i y_4$.  We will  also take
the metric to be the most symmetric possibility, with:
\begin{equation}\label{simpmet} 
A^{r}_{ij}=A^{y}_{ij}=\delta_{ij}\, ,\ \  A^{m}_{ij}=0 \,.
\end{equation}
The equations of motion then reduce to:
\begin{equation}\label{Einred}  
 f^2 ~+~  \coeff{1}{4} \, \beta^2  |b|^2 ~=~ 1\, . 
\end{equation}

The pp-wave limit with maximal supersymmetry satisfies~\eqref{Einred} with
$f= \pm 1, b=0$.   Recall that this is the pp-wave limit of both
$AdS_5\times S^5$  
and $AdS_5\times T^{1,1}$.  The solution we present below as the
Penrose limit of  
the gravity dual of the  Leigh-Strassler fixed point is also of the simplified
form~\eqref{simpmet}, satisfying \eqref{Einred}, but with: 
$$ f= - {1\over 2}\, ,\qquad b= i\, \sqrt{3}\,, \qquad \beta=1\,. $$ 
The Ansatz \eqref{ppfluxsol}, \eqref{specB} and \eqref{simpmet} with 
parameters satisfying \eqref{Einred}, defines a
2-sphere of solutions that interpolate between the solution of
defined below and the point of maximal supersymmetry.  Thus this
interpolating family is related to the Penrose limit of the  family
of solutions which interpolate between the  orbifold of the
solution of
\cite{PilchEJ}  and conifold fixed points in the supergravity duals of 
$\CN=2$ quiver gauge theories.  

\subsection{The $\cN=2$ supersymmetric $AdS$ solution}

Here we summarize the essential features of
 the $\cN=2$ supersymmetric supergravity dual of the $\CN=1$
``Leigh-Strassler'' field theory fixed point in four dimensions.  The
five-dimensional gauged supergravity solution was found in \cite{KhavaevFB}, 
and this was subsequently ``lifted'' to ten-dimensional, IIB supergravity
in~\cite{PilchEJ}. The ten-dimensional metric takes the form:
\begin{equation}\label{metrten} 
ds_{10}^2\eql \Omega^2 \,ds^2_{{\rm AdS}_5} - ds_5^2\, . 
\end{equation}
For the $AdS_5$ directions, we take 
\begin{equation}\label{adsmetr}
ds^2_{{\rm AdS}_5}\eql L^2 \left( 
\cosh^2\rho\, dt^2  -  d\rho^2 - \sinh^2\rho\, d\Omega_3^2
\right)\, , 
\end{equation}
where  $d\Omega_3^2$  is the metric on the
unit $3$-sphere.  The parameter, $L$, is the ``radius'' of the
$AdS_5$.  The internal metric is:
\begin{equation}\label{themetric}
\begin{split}
 ds_5^2&\eql {\sqrt{3}\over 8}a^2 (3-\cos(2\theta))^{1/2}\left(
 d\theta^2 +{\cos^2\theta\over
 3-\cos(2\theta)}((\sigma_1)^2+(\sigma_2)^2) \right. \\
& \hskip4.4cm \left. 
+{\sin^2(2\theta)\over(3-\cos(2\theta))^2}(\sigma_3)^2  \right)\\
 &\qquad + {\sqrt{3}\over 12}a^2(3-\cos(2\theta))^{1/2}\left(d\phi+
 {2\cos^2\theta\over 3-\cos(2\theta)} \sigma_3\right)^2\,,
\end{split}
\end{equation}
where the scale, $a$, is given by:
$$a = {2^{13/6} \over 3} L\,, $$
and the warp factor $\Omega$ in~\eqref{metrten} is
\begin{equation}\label{wrnow}
\Omega^2\eql 2^{1/3}\left(1-{1\over 3}\cos(2\theta)\right)^{1/2}\,.
\end{equation}
This solution also has a 5-form flux:
\begin{equation}\label{fiveform}
F= - {2^{5/3} \over 3 L \Omega^5} ( 
e^1\wedge \cdots \wedge e^5 +  e^6\wedge \cdots \wedge e^{10} )\,,
\end{equation} 
and a 3-form flux which can be obtained from the complex 2-form potential
\begin{equation}\label{exacttwo}
\cB = - {i \over \sqrt{3}}\, e^{-2i \phi} \,
(e^6 + i \, e^9)\wedge (e^7-i\, e^8). 
\end{equation}
In equation~\eqref{fiveform}, we have included a factor of 2 
to correct a typo in~\cite{PilchEJ} \footnote{The correct coefficient
can be obtained from  
subsequent papers: \cite{PilchFU,KhavaevYG}.}.

It will be useful later to recall some further details of \cite{PilchEJ}.
First, one can parametrize $\IR^4$ using an $SU(2)$ transformation
based upon Euler angles, $\varphi_j$:
\begin{equation}\label{zetadefn}\begin{split}
& \zeta_1 = y_1 + i\, y_2 =  y\, \cos(\coeff{1}{2} \, \varphi_1)\, 
e^{- {i \over 2}\, (\varphi_3 + \, \varphi_2)} \,,  \\
& \zeta_2 =  y_3 + i\, y_4 =  y\, \sin(\coeff{1}{2} \, \varphi_1)\, 
e^{-{i \over 2} \, (\varphi_3 - \, \varphi_2)} \,. 
\end{split}\end{equation}
The left-invariant one-forms, $\sigma_j$, can then be written
explicitly as:
\begin{equation}\label{sigdefn}
\sigma_1 ~\pm~ i\, \sigma_2 ~=~ 
e^{\pm i\, \varphi_3} \, (d\varphi_1 \mp  i\, \sin(\varphi_1) \,
d\varphi_2)  \,, \qquad 
\sigma_3 ~=~ d\varphi_3 + \cos(\varphi_1)\, d\varphi_2 \,. 
\end{equation}
In \cite{PilchEJ} the $S^5$ was parametrized by taking:
\begin{equation}\label{ucoords}
u_1 = \zeta_1\, e^{-i \, \phi/2}\,, \quad 
u_2 = \zeta_2\, e^{-i \, \phi/2}\,, \quad u_3 = \sin(\theta) \, e^{-i
\, \phi }\,. \end{equation} 
with $y= \cos(\theta)$ so that $|u_1|^2 + |u_2|^2 +|u_3|^2  =1$.
Since the metric and the tensor gauge fields are  constructed using the
$\sigma_j$ there is a manifest   $SU(2)$ symmetry.  The metric
also has a $U(1) \times U(1)$ symmetry under $\varphi_3$ and 
$\phi$ translations, but the $B$-field \eqref{exacttwo} breaks this to
\begin{equation}\label{Rsymm}
\phi \to \phi - \alpha\,, \qquad 
\varphi_3 \to \varphi_3 + 2 \alpha\,,
\end{equation}
since $ (e^7-i\, e^8) \sim (\sigma_1 - i \sigma_2) \sim 
e^{-i \, \varphi_3}$.
The residual ($\cN=1$) supersymmetry generators on the
brane transform under this as
$\epsilon_\pm \to e^{\pm i \alpha} \epsilon_\pm$,
and so \eqref{Rsymm} represents the canonically normalized
$\cR$-symmetry of the  $\cN=1$ superconformal theory
on the brane.

It is simple to modify the foregoing solution in order to describe an analogous
$\CN\! =\! 1$ fixed point in $\CN\! =\! 2$ $A_{n-1}$ quiver gauge
theory. The UV 
dual is IIB on AdS$_5\times S^5/\BZ_n$, where the orbifold action
identifies the Euler angle $\varphi_3\sim \varphi_3 + 4\pi /n$. The
$\CN=1$ orbifold IR fixed point is described by the above solution
together with this identification on $\varphi_3$.

\subsection{The Penrose limit}

After substituting all the factors into \eqref{themetric} one finds
terms of the  
form:
\begin{equation}\label{metpart}
 L^2 \, \Omega^2\,\Big(  \cosh \rho \, dt^2 ~-~ {4 \over 9} \,
\Big(d\phi+  {2\cos^2\theta\over 3-\cos(2\theta)} \sigma_3\Big)^2 ~+~ 
\dots \Big) \,. 
\end{equation}
We thus consider  geodesics that lie  near $\rho = 0$, $\theta
=\pi/2$, and have $x^- \sim L^2(t +{2 \over 3} \phi)$.  
It is convenient to introduce some additional constants
to clean up the result, and so we make the coordinate redefinitions:
\begin{equation}\label{newcoords}
t= x^+,~~\rho = {3^{1/4}\over 2^{2/3}} \, {r\over L},~~
\theta = {\pi \over 2} - {3^{3/4}\over 2^{7/6}}\, {y\over L},~~
\phi = {3\over 2} 
\left({\sqrt{3}\over 2^{4/3}}\, {x^-\over L^2} - x^+\right), 
\end{equation}
In the limit $L\rightarrow \infty$ we find the metric:
\begin{equation}\label{ppwaveform}
 ds^2 =  2\, dx^- \, dx^+ + (r^2+y^2) (dx^+)^2
-  d\vec{r}^{\, 2}  - ds_4^2,
\end{equation} 
where 
\begin{equation}\label{cartesians}
\begin{split}
d\vec{r}^{\, 2}  ~=~  & dr^2 ~+~ \coeff{1}{4} \, r^2   \, d\Omega_3^2, \\
ds_4^2 ~=~ & dy^2  ~+~  \coeff{1}{4} \,y^2  \,  \left( (\sigma_1)^2 
+(\sigma_2)^2 +(\sigma_3 - 2\, dx^+)^2 \right) \,.
\end{split}\end{equation}

The same limit of the 5-form  yields:
\begin{equation}\label{ppfiveform}
F =  - {1 \over 2} \, dx^+\wedge 
( dr_1\wedge\cdots \wedge dr_4 + dy_1\wedge\cdots \wedge dy_4 ). 
\end{equation} 
while the complex 2-form potential~\eqref{exacttwo}  becomes
\begin{equation}\label{twoform}
\cB = \coeff{i }{ 2 \, \sqrt{3}}\, e^{3\, i x^+} \, y\,
 ( dy - \coeff{i}{2}\, y\, \sigma_3)\wedge (\sigma_1 -i \, \sigma_2) ~=~
- \coeff{i }{  \sqrt{3}}\, e^{3\, i x^+}  \,d\zeta_1 \wedge d\zeta_2
 \,. 
\end{equation}

There are now two natural changes of variable:  (i)  $ \varphi_3
\to \varphi_3 + 2x^+$, and  (ii)  $ \varphi_3 \to \varphi_3 + 3x^+$.
The former removes the cross-term in the metric \eqref{cartesians},
while the latter removes the  $x^+$ dependence in $\cB$.  
This removal of the $x^+$-dependence is an important step
in computing the normal modes of the string, and so we will 
return to it in section 4.

To get to the Ansatz  \eqref{specB} one first makes the shift: 
$ \varphi_3 \to \varphi_3 + 2x^+$.  This generates an extra term in
$\cB$ that is proportional to $dx^+ \wedge (\sigma_1 -i \, \sigma_2)$
and which may be removed by adding a pure gauge term, $d\cA$,
where:
\begin{equation}\label{puregauge}
\cA ~=~ \coeff{i }{\sqrt{3}}\, e^{  i x^+} \, y^2 \,
 (\sigma_1 -i \, \sigma_2) \,. 
\end{equation}
The final result is: 
\begin{equation}\label{Bmetfnl}\begin{split}
ds_4^2 ~=&~ d\vec y^{\, 2}  ~=~   dy^2 +  \coeff{1}{4} \,y^2  \,  \big( 
(\sigma_1)^2  +(\sigma_2)^2 +(\sigma_3 )^2 \big) \,, \\
\cB ~=&~ i\, \sqrt{3} \, e^{i x^+} \, d\zeta_1 \wedge d\zeta_2\,. 
\end{split}\end{equation}
Thus the  pp-wave metric is the simplest possible form: 
on the transverse coordinates  $\BR_r^4\oplus \BR_y^4$ the symmetric 
matrix, $A$, in \eqref{ppfluxsol} is the identity matrix.  It fits the
Ansatz above  
with $\beta = 1, f = -{1\over 2}$ and $b =i \sqrt{3}$.  Note
that $\beta = - 2f$.

Finally, note  that for the orbifold, $S^5/\Gamma$, the $\varphi_3$
coordinate of $\BR_y^4$ is still 
identified under the orbifold action. So after taking the scaling limit of  
AdS$_5\times S^5/\Gamma$, the space of the $\vec{y}$ is  
$\BR_y^4/\Gamma$, for the choice of geodesic in~\eqref{newcoords}.

\section{Supersymmetry and Killing Spinors} 

For vanishing dilaton and axion, the supersymmetry conditions are
\begin{equation}\label{gravitino}\begin{split}
& D_M\, \epsilon
+{i\over 480} F_{PQRST}\, {\gamma^{PQRST}} \gamma_M\, \epsilon
+{1\over 96} G_{PQR}\, ({\gamma_M}^{PQR} \\
& \hskip7cm -9\, {\delta_M}^P\, \gamma^{QR})\, \epsilon^* = 0,
\end{split}\end{equation}
and
\begin{equation}\label{dilatino}
\gamma^{MNP}\, G_{MNP}\,\epsilon =0. \end{equation}

\subsection{Integrability conditions}

The easiest way to count the number of supersymmetries  is to 
compute the integrability, or zero-curvature condition for the differential
operator in \eqref{gravitino}.  The result of doing this is an
algebraic condition 
of the form:
\begin{equation}\label{intcond}
\begin{pmatrix}
A & B \\ C & D 
\end{pmatrix} 
\begin{pmatrix}
\epsilon \\ \epsilon^*\end{pmatrix} 
 ~=~ \begin{pmatrix}0 \\ 0 \end{pmatrix} \,. \end{equation}
We consider a metric of the form given in  \eqref{ppfluxsol} and
 \eqref{simpmet},  
 and we take $F$ to be as in \eqref{ppfluxsol}, with $\cB$ given by
\eqref{specB}.  We recall that the only non-trivial field equation is 
\eqref{Einred}. For this background it  is straightforward to solve
 \eqref{intcond} 
and \eqref{dilatino}. 

Introduce frames  $e^A$ for the metric in \eqref{ppfluxsol}\ with 
$e^j = dr_j, e^{j +4} = dy_j, j=1,\dots,4$, and with $e^0$ and $e^9$
as the remaining time-like and space-like components.  Define
$\gamma^\pm  = {1 \over \sqrt{2}} (\gamma^0 \pm  \gamma^9)$.
As is by now familiar, there are always 16 solutions to these conditions
given by the zeroes of:
\begin{equation}\label{genericsol} \gamma^+ \, \epsilon ~=~  \gamma^+
\, \epsilon^* ~=~ 0\,.  
\end{equation}
One finds four more non-trivial solutions to  \eqref{dilatino}  and
\eqref{intcond}
{\it if and only if} $\beta = - 2f$.  The solutions are present for
all values  of $b$ and $f$ satisfying \eqref{Einred}, and the
solutions are precisely 
those that solve the projection conditions:
\begin{equation}\label{projconds} 
\gamma^+(\gamma^5 + i\, \gamma^6) \, \epsilon ~=~
\gamma^+(\gamma^7 + i\, \gamma^8) \, \epsilon ~=~ 0\,. 
\end{equation}

The integrability conditions are necessary conditions for supersymmetry,
and are usually, but not always, sufficient.  To arrive at a
sufficient condition 
one must check higher order integrability using further commutators of
the supercovariant derivative \cite{vanNieuwenhuizenWU}.  For the background 
we are considering here, it turns out that these extra integrability conditions
are trivially satisfied.   Alternatively, one can see that there must
be at least 
four additional supersymmetries by boosting the solutions of \cite{PilchEJ}.

\section{The modes of the Green-Schwarz string}

It is convenient to introduce a scale $\mu$ by rescaling
$x^-\rightarrow x^-/\mu$, $x^+\rightarrow \mu x^+$. This changes the
metric to:
\begin{equation}\label{rescaledmet} 
ds^2 = 2\, dx^- \, dx^+  + \mu^2(r^2+y^2) (dx^+)^2
- d\vec{r}^{\, 2}  -  d\vec{y}^{\, 2}   \, , 
\end{equation} 
and it also introduces factors of $\mu$ into $F$ and $\cB$.

\subsection{The bosonic modes} 

Other than the metric, the only  background field that couples to the
bosonic string is  the NS-NS two-form, which is is the real part of
$\cB$.  After 
the re-scaling, this is given by
\begin{equation}\label{rescaledB}
 B =   {1  \over 2}\, \left(
b\, e^{i \beta \mu x^+} \, d\zeta_1 \wedge d\zeta_2
+ \bar{b}\, e^{-i \beta \mu x^+} \, 
d\bar{\zeta}_1 \wedge d\bar{\zeta}_2 \right) \, .
\end{equation}
The bosonic part of the Green-Schwarz Lagrangian is:
\begin{equation}\label{gsbosact} 
\CL =  \coeff{1}{2} \, g_{\mu\nu} \,
\partial_\alpha 
x^\mu\, \partial^\alpha x^\nu
+ B_{\mu\nu} \,\partial_\tau x^\mu \partial_\sigma x^\nu\,,
\end{equation}
In the light-cone gauge, one sets $x^+=\tau$,  and then
\eqref{gsbosact} becomes
\begin{equation}\label{gslc}
\begin{split}
\CL = &- {1\over 2} (\partial_\alpha r_i)^2 
- {1\over 2}|\partial_\alpha \zeta_I|^2
- {\mu^2\over 2} \left( (r_i)^2 + |\zeta_I|^2 \right)\\
& +  {\mu \over 2}\, \left(
b\, e^{i\beta \mu \tau} \, 
(\partial_\tau \zeta_1  \,\partial_\sigma \zeta_2
-\partial_\sigma \zeta_1  \,\partial_\tau \zeta_2 ) \right. \\
& \left. \hskip2cm 
+ \bar{b}\, e^{-i\beta\mu \tau} \, (\partial_\tau \bar{\zeta}_1 \,
\partial_\sigma \bar{\zeta}_2 
-\partial_\sigma \bar{\zeta}_1 
 \,\partial_\tau \bar{\zeta}_2 )\right),
\end{split}\end{equation}

The four bosons, $r_i$, form a set of four oscillator towers 
with the same spectrum as in the pp-wave background without
any 3-form flux:
\begin{equation}\label{ppfreq}
\omega_0^2 =\mu^2 + {n^2\over (\alpha' p^+)^2} . 
\end{equation}
The $\zeta_I$ system has a set of coupled equations of motion
\begin{equation}\label{zetaeom}\begin{split}
& (-\partial_\tau^2 +\partial_\sigma^2) \zeta_1 - \mu^2 \,\zeta_1
- \bar{b}\mu  \left(
\partial_\tau (e^{-i\beta\mu\tau}\, \partial_\sigma \bar{\zeta}_2)
- e^{-i\beta\mu\tau}\, \partial_\tau\partial_\sigma \bar{\zeta}_2\right) =0,\\
& (-\partial_\tau^2 +\partial_\sigma^2) \bar{\zeta}_2 - \mu^2 \,\bar{\zeta}_2
+  b\mu  \left(
\partial_\tau (e^{i\beta\mu\tau}\, \partial_\sigma \zeta_1)
- e^{i\beta\mu\tau}\, \partial_\tau\partial_\sigma \zeta_1\right)=0, \\
\end{split}\end{equation}
and their complex conjugates.   

The equations \eqref{zetaeom} are separable, and so 
we can expand into Fourier modes in the $\sigma$ coordinate:
\begin{equation}\label{oscill}\begin{split}
& \zeta_1 = \sum_{n\geq 0} \alpha^{(1)}_n(\tau) \,
e^{-in\sigma/ (\alpha' p^+) - i(\beta\mu\tau+{\rm arg}(b))/2}, \\
& \zeta_2 = \sum_{n\geq 0} \alpha^{(2)}_n(\tau) \,
e^{in\sigma/ (\alpha' p^+) - i(\beta\mu\tau+{\rm arg}(b))/2}\,, \\
\end{split}\end{equation}
and we then find   coupled ODE's for the $\alpha^{(k)}_n(\tau) $:
\begin{equation}\label{odes}\begin{split}
& \ddot{\alpha}^{(1)}_n -  i\beta \mu\,  \dot{\alpha}^{(1)}_n
+ \left(\omega_0^2 - {\beta^2 \mu^2 \over 4} \right) \alpha^{(1)}_n 
- {n \beta|b| \mu^2 \over 2 \alpha' p^+} \,
 \bar{\alpha}^{(2)}   =0, \\
& \ddot{\bar{\alpha}}^{(2)}_n +  i\beta \mu\,  \dot{\bar{\alpha}}^{(2)}_n
+ \left( \omega_0^2- {\beta^2\mu^2 \over 4} \right)
\bar{\alpha}^{(2)}_n
- { n \beta |b| \mu^2\over 2\alpha' p^+} \,\alpha^{(1)}   =0,
\end{split}\end{equation}
where $\omega_0$ is given by~\eqref{ppfreq}. Note that the shifts by
$\exp(\pm {i \over 2} \beta\mu\tau) $ in \eqref{oscill}  were used to remove 
the factors  of $e^{i\beta\mu\tau}$  from the system.  This was possible as
a  consequence of the special form \eqref{specB} of $\cB$.  These shifts also
correspond to sending $\varphi_3 \to \varphi_3  + \beta x^+$, which removes
the $x^+$-dependence from $\cB$.

Finally, we   obtain the frequencies of the string modes from the
eigenvalues of this linear system:
\begin{equation}\label{freq}
\omega_\pm^2 =  \left[\omega_0^2   +\left( {\beta\, \mu\over 2}\right)^2  
\pm  \beta \,\mu \, 
\sqrt{\omega_0^2 + \left( {n\, |b| \over 2\, \alpha' p^+ }\right)^2  }
\right].
\end{equation}

It is interesting to observe that at least this part of the string spectrum
is not invariant under the $SU(2)$ symmetry of the family of
supergravity solutions parametrized by $b$ and $f$.  Indeed, one 
could have seen this {\it ab initio} because the bosonic string action
does not depend upon $F_{(5)}^{RR}$ and so cannot depend 
directly upon $f$:  it is thus impossible for the $SU(2)$ invariant
combination \eqref{Einred} to appear in the spectrum\footnote{It might be 
possible for $f$ to appear in these equations
via the supersymmetry condition $\beta = -2f$, but even with
this, \eqref{freq} is not $SU(2)$ invariant. }.  So string theory does indeed 
break the $SU(2)$, confirming that it can only be a large $N$ symmetry of the
field theory on the brane.  As one should also expect, the particle-like
excitations with $n=0$ are independent of both $f$ and $b$, and thus
respect the $SU(2)$ symmetry.

\subsection{The fermionic modes}
 
There are several expressions of the quadratic fermionic part of the Green-Schwarz
action in a general background \cite{CveticSI,CveticZS, 
MetsaevRE, RussoRQ}, but sadly there are several inconsistencies in
signs, factors and normalizations.   Fortunately,
there is one clear principle that defines the relevant part of the
action \cite{MetsaevRE}: 
The differential operator is precisely the supercovariant derivative that appears
in the gravitino variation.  To be more precise, the part of the Green-Schwarz action
that is quadratic in fermions is \cite{MetsaevRE}:
\begin{equation}\label{fermact}  
\cL_{2F} ~=~ i (\eta^{ab} \delta_{\cI \cJ} -  \epsilon^{ab} \, 
\rho_{3\, \cI\cJ})\, \partial_a x^{M }\  \bar \theta^\cI \gamma_{M} \cD_b
\theta^\cJ  \   \,,
\end{equation}
where $\cD_a$ is the pull-back of the supercovariant derivative:
\begin{equation}\label{scderiv}
\begin{split}\cD_a  ~\equiv~   \partial_a 
  ~+~ \partial_a x^M \Big[ ~ 
\big(& \coeff{1}{4}\,  \omega_{M\, A B}  - \coeff{1}{8}\, H_{M\, AB}
 \, \rho_3\big) 
 \gamma^{ AB}  \\ ~-~ 
&\coeff{1}{48}\, F_{ABC} \, \gamma^{ABC}\,  \rho_1 \,\gamma_M \\ 
& +
\coeff{1}{480}\, F_{ABCDE} \, \gamma^{ABCDE}\, \rho_0 \, \gamma_M~\Big]
\,. 
\end{split}\end{equation}
The matrices, $\rho_j$, are defined in \cite{MetsaevRE}, and following 
this reference, we take $\eta^{00} = - \eta^{11} =-1$, 
$\epsilon^{01} = 1$.  


While it is not immediately obvious, \eqref{scderiv}  agrees with the results
of~\cite{MetsaevRE}.  The  $5$-form in  \cite{MetsaevRE} is twice that of
\cite{SchwarzQR}
and hence twice that of this paper.  There is also an apparent
discrepancy
of a factor of two in the normalization of the $RR$ $3$-form, however,
this
normalization depends upon a choice of the constant value of the
dilaton.    In this paper we are taking the dilaton to be
zero ($e^{\phi} =1$), whereas\footnote{We are grateful to A.~Tseytlin for 
clarifying this.}  \cite{MetsaevRE}  takes $e^{\phi} =2$.
We have thus adjusted the  normalizations in \eqref{scderiv} so as
to be consistent with our conventions for the $RR$ forms.
One can confirm that  \eqref{scderiv} is properly normalized by checking that
it
is consistent with the gravitino variation, \eqref{gravitino} of
\cite{SchwarzQR}. 


The first step is to pass between the real and complex bases by
taking $\epsilon = \theta^1 + i \theta^2$, and writing
$G_{(3)} = H_{(3)}+ i F_{(3)}$.  Beyond this, 
it seems that \eqref{scderiv} and \eqref{gravitino} are very different:
The former manifestly preserves an $SU(1,1)$ duality group and, in
particular, preserves the $U(1)$ that rotates $H_{(3)}$ into $F_{(3)}$.
The latter formula apparently breaks this symmetry.  The key to
understanding the difference is simply that the formul\ae of
\cite{SchwarzQR} are all in the Einstein frame, whereas the string
actions must, 
of course, be in the string frame.  The passage to the string frame 
also involves mixing the gravitino with the dilatino (see, for example,
Appendix B of \cite{HassanBV}): 
$$
\psi_M ~\to~ \psi_M  ~-~ {i \over 4} \, \gamma_M \, \lambda \,.
$$
(There is a sign difference here compared to \cite{HassanBV} because
of our different $\gamma$-matrix conventions.)
This means that to pass to the string frame one must add a multiple
of \eqref{dilatino} to \eqref{gravitino}.  If one does this then one
does, indeed, arrive 
at \eqref{scderiv}.  

For the particular class of background that we are considering here one
obtains the following:
\begin{equation}\label{fermderiv}\begin{split}
 \cD_a \begin{pmatrix} \theta^1 \\ \theta^2 \end{pmatrix}  ~=~  &
\partial_a \begin{pmatrix}\theta^1 \\ \theta^2\end{pmatrix} \\
  ~&+~
(\partial_a x^+)\,  
\Big[   \coeff{1}{4}\,  \omega_{+\, A B} \, \gamma^{ AB}  \,
\begin{pmatrix} \theta^1   
\\ \theta^2\end{pmatrix}  - \coeff{1}{8}\, H_{+ \, AB}   \,  \gamma^{ AB} \, 
\begin{pmatrix} \theta^1 \\ - \theta^2\end{pmatrix}    \\ & -   \coeff{1}{8}\, 
F_{+ \, AB}   \,  \gamma^{ AB} \, \begin{pmatrix}\theta^2 \\
\theta^1\end{pmatrix}    
  -  \coeff{f}{2}\,  \big( \gamma^{1234} +\gamma^{5678}  \big) \,
\begin{pmatrix} -\theta^2 \\   \theta^1\end{pmatrix}   ~\Big]
\,.  
\end{split}\end{equation}
In spite of the string frame, the special form of the background preserves
the $U(1)$ symmetry between $H_{(3)}$ and $F_{(3)}$.  Using this, the
fermion equations reduce to:
\begin{equation}\label{fermeqn}\begin{split}
\gamma_+\begin{pmatrix}(\partial_{\tau} + \partial_\sigma)\, \theta^1 \\
(\partial_{\tau} -  
\partial_\sigma)\, \theta^2 \end{pmatrix}
= &+\coeff{1}{4}\, \gamma_+ \not \!\! H_{(3)} 
\begin{pmatrix}\theta^1  \\ - \theta^2\end{pmatrix}
+ \coeff{1}{4}\, \gamma_+ \not \!\! F_{(3)} 
\begin{pmatrix}\theta^2  \\  \theta^1\end{pmatrix}\\
& +\coeff{1}{2}\, \gamma_+ \not \!\! F_{(5)} 
\begin{pmatrix}\theta^2  \\ - \theta^1\end{pmatrix} \,,
\end{split}\end{equation}
where 
\begin{equation}\label{slashdefns}
\not \!\! H_{(3)}= \coeff{1}{2}\, H_{+\mu\nu}\gamma^{\mu\nu}, \ \ 
\not \!\! F_{(3)}=\coeff{1}{2}\, F_{+\mu\nu}\gamma^{\mu\nu}, \ \ 
\not \!\! F_{(5)}=\coeff{1}{24}\, F_{+\mu\nu\sigma\lambda}
\gamma^{\mu\nu\sigma\lambda}\,. 
\end{equation}

As with the bosonic equations, the fermionic equations are separable. 
The detailed solution may be found in Appendix A, but we will summarize the
key points here.

Half of the modes do not couple to the background $B$-field and so the mode
analysis is elementary.   We thus find eight sets of fermionic
oscillators with  
frequencies:
\begin{equation}\label{fermfreqone}
\omega ~=~ \pm \, \sqrt{(f\, \mu)^2 ~+~ 
\Big({n \over \alpha' p^+}\Big)^2} \,,
\end{equation}
with each sign having multiplicity $4$.     The remaining fermions 
couple to the $B$-field and so we have to shift them by phases,
exp$(\pm i(\beta\mu\tau+arg(b))/2)$,  much as we did for the bosons.
We then obtain two copies of the following  first order system of  linear 
differential equations:
\begin{equation}\label{fermsys}\begin{split}
& \Big(\partial_\tau \mp \coeff{1}{2}\, i\beta\mu + \Big({i n \over \alpha'
p^+}\Big)  \Big)\,  
\begin{pmatrix}
\alpha_{(n)}  \\ \widetilde \alpha_{(n)} \end{pmatrix}
\\
& \hskip3cm 
=- \coeff{1}{ 2}\, |b|\beta \mu\, \begin{pmatrix}i \widetilde \alpha_{(n)}  -
  \widetilde \beta_{(n)}  \\ i\alpha_{(n)}  + \beta_{(n)} \end{pmatrix}
- f \, \mu\, \begin{pmatrix}\beta_{(n)}  \\  \widetilde
\beta_{(n)}\end{pmatrix} \,, \\  
& \Big(\partial_\tau \mp\coeff{1}{2}\, i\beta\mu -\Big({i n \over \alpha'
p^+}\Big)\Big)\,  
\begin{pmatrix}\beta_{(n)}  \\  \widetilde \beta_{(n)} \end{pmatrix}  
 \\
& \hskip3cm
=\coeff{1}{ 2}\, |b|\, \beta \mu\, \begin{pmatrix}i  \widetilde \beta_{(n)}  +
 \widetilde \alpha_n  \\ i\beta_{(n)} - \alpha_{(n)} \end{pmatrix}
+f \,\mu\, \begin{pmatrix}\alpha_{(n)} \\  \widetilde \alpha_{(n)}
 \end{pmatrix} \,. 
\end{split}\end{equation}
The normal modes of this system are:
\begin{equation}\label{fermmodes}\begin{split}
&\pm \left[ \bigg(\coeff{1}{2}\,  |b|^2\,\beta^2 \mu^2+f^2\mu^2+
p^2 +
\coeff{1}{4}\, \beta^2 \mu^2 \bigg)\right. \\
& \hskip2cm\left. \pm \coeff{1}{2}\,  {\beta \mu}\,
\sqrt{( 2f- |b|^2\beta)^2 \mu^2 + 4\, p^2(1+|b|^2)}\right]^{1/2}
\end{split}\end{equation}
with all four permutations of the $\pm$ signs, and where $p  \equiv
 {n \over \alpha' p^+}$.
We thus find eight fermions having each of these frequencies with multiplicity
two.   

If one uses the equations of motion, \eqref{Einred},  and the supersymmetry
condition $\beta = - 2f$, then these frequencies  reduce to the much 
simpler form:
\begin{equation}\label{fermfreqthree}
\omega ~=~ \pm\, \mu^2 \pm  \sqrt{ \mu^2 + 
\Big({n \over \alpha' p^+}\Big)^2 }\,,
\end{equation}
with all four permutations of the $\pm$ signs, each with a multiplicity
of two.

It is interesting to note that the  modes \eqref{fermfreqthree} do not
depend upon 
the values of $f$ and $b$, whereas the modes \eqref{fermfreqone} do,
and thus 
this part of the string spectrum is not $SU(2)$ invariant.

\subsection{Frequencies, energy and charge}

Having found the eigenmodes of the string excitations, we would like to 
relate the frequencies to the physical quantum numbers, and most 
particularly, we would like to know how this works for the Penrose
limit considered in section 2.

To find the normal modes we had to make shifts by
$\exp(\pm{i \over2} \beta \mu \tau)$, and this corresponds to 
going to a coordinate system in which the $\cB$ field is
independent of $x^+$.   For the general family considered
in section 2, this means redefining:
\begin{equation}\label{newvphi} 
\varphi_3 = \tilde \varphi_3+ \beta x^+ \,,
\end{equation}
Thus time translations in our systems of differential equations
are associated to a combination of $x^+ \to x^+ + \alpha$ and
$\varphi_3  \to \varphi_3  +\beta \alpha$
(so that $\tilde \varphi_3$ remains fixed).  Thus the frequencies, $\omega$, 
computed above are associated to the corresponding mix of N\"other charges.

In the AdS/CFT correspondence one should recall that time
translations, $t \to t + \alpha$   yield  
a N\"other charge that is the quantum number, $\Delta$, representing
the conformal dimension of the operator on the brane.  In \cite{PilchEJ}
the $U(1)_\cR$ symmetry was identified as \eqref{Rsymm}, at least for
$\beta=1$. 
In taking the Penrose limit we had to shift  $\varphi_3  \to \varphi_3
+ 2 x^+$, 
and to diagonalize the system of eigenmodes we had to make a further
shift to the $x^+$-independent configuration.   We are thus led to
the following coordinates that describe the  $x^+$-independent configuration:
\begin{equation}\label{newcoordstwo} 
x^+ = t \,, \qquad x^- \sim L^2 \,
(t + \coeff{2}{3} \, \phi)\,, 
\qquad \hat \varphi_3 = \varphi_3  - 3\,t \,. 
\end{equation}
Shifting $t \to t + \alpha$ and keeping the new coordinates fixed requires
$\varphi_3  \to \varphi_3 + 3\alpha$, $\phi   \to \phi - {3 \over 2} \alpha$.
Hence the time translations  in our system of differential equations
is associated to the conserved quantity:
$$
\Delta - {3 \over 2}  \, \cJ\,.
$$

\section{Final Comments}

We have found a new family of pp-wave solutions of IIB supergravity.
The family has constant $5$-form flux, non-constant $3$-form
fluxes, and time varying dilaton and axion.  There are obviously many 
generalizations of the solution presented here, both within IIB supergravity
and in $\cM$-theory.  We have not attempted to classify the
broad class of possibilities, but instead we have focused on the solutions
that are most closely related to interesting results within the AdS/CFT correspondence.
Even within this context there are more general possibilities: one
can presumably find Penrose limits of any of the flow solutions.

There are several important features of our solutions:  first,  the
string theory is exactly solvable in these backgrounds.  Secondly,
our solutions are Penrose limits of  AdS solutions that
are interesting as holographic duals of field theory fixed points.
Finally,  our backgrounds {\it are not} maximally supersymmetric.
Thus our results enable one to perform some deeper ``stringy''
 tests of the holographic  duals of $\cN=1$ supersymmetric flows.

\bigskip
\leftline{\bf Acknowledgments} 
This work was supported in part by funds  
provided by the DOE under grant number DE-FG03-84ER-40168.
NH is supported by a Fletcher Jones Graduate Fellowship.
We would like to thank D.\ Berenstein and J.\ Gomis for discussions
and are particularly grateful A.\ Tseytlin for helpful correspondence.

\begin{appendix}
\section{Fermionic Spectrum}

The equations of motion for the fermion system are:

\begin{equation}\label{eomthree}\begin{split}
\gamma_+ \begin{pmatrix} (\partial_{\tau} + \partial_\sigma)\, \theta^1 \\
(\partial_{\tau} -  
\partial_\sigma)\, \theta^2 \end{pmatrix}
= & +\coeff{1}{4}\, \gamma_+ \not \!\! H_{(3)} 
\begin{pmatrix}\theta^1  \\ - \theta^2\end{pmatrix}
+ \coeff{1}{4}\, \gamma_+ \not \!\! F_{(3)} 
\begin{pmatrix}\theta^2  \\  \theta^1\end{pmatrix} \\
& +\coeff{1}{2}\, \gamma_+ \not \!\! F_{(5)} 
\begin{pmatrix}\theta^2  \\ - \theta^1\end{pmatrix} \,,
\end{split}\end{equation}
where 
\begin{equation}\label{slash}
\not \!\! H_{(3)}= \coeff{1}{2}\, H_{+\mu\nu}\gamma^{\mu\nu}, \ \ 
\not \!\! F_{(3)}=\coeff{1}{2}\, F_{+\mu\nu}\gamma^{\mu\nu}, \ \ 
\not \!\! F_{(5)}=\coeff{1}{24}\, F_{+\mu\nu\sigma\lambda}
\gamma^{\mu\nu\sigma\lambda}. 
\end{equation}
One could solve this by using the explicit form of the spacetime Dirac
matrices,  
or as we will, proceed by realizing these matrices as creation and destruction 
operators. 
\begin{equation}\label{anncreat}
a_j^\dagger = {i \over 2}(\gamma^{2j-1} + i\gamma^{2j }) \,, 
\quad a_j = {i \over 2}(\gamma^{2j-1} - i\gamma^{2j }) \,,  \qquad
j=1,\dots,4 \,. 
\end{equation}
Note that these operators have a factor of $i$ in front of
them since we are using anti-hermitian $\gamma$-matrices
as in \cite{SchwarzQR}.
For ${\bf 8_s}$ we take the spinor basis to be:
\begin{equation}\label{explicitbasis}
\begin{pmatrix}|0\rangle  \\ a^\dagger_1 a^\dagger_2|0\rangle  
\\ a^\dagger_1 a^\dagger_3|0\rangle  
\\ a^\dagger_1 a^\dagger_4|0\rangle  \\ a^\dagger_2 a^\dagger_3|0\rangle  
\\ a^\dagger_2 a^\dagger_4|0\rangle  \\ a^\dagger_3 a^\dagger_4|0\rangle  
\\ a^\dagger_1 a^\dagger_2 a^\dagger_3 a^\dagger_4|0\rangle \end{pmatrix}
\end{equation}
We can now express the matrices in \slash\ as:
\begin{equation}\label{eomfour}\begin{split}
&\not \!\! H_{(3)} = -2ib\beta\mu\, e^{i\beta\mu\tau} a^\dagger_3 a^\dagger_4
+ 2i{\bar b}\beta\mu \, e^{-i\beta\mu\tau}a_3 a_4 \\
&\not \!\! F_{(3)} = - 2b\beta\mu\, e^{i\beta\mu\tau}a^\dagger_3 a^\dagger_4
- 2{\bar b}\beta\mu\, e^{-i\beta\mu\tau}a_3 a_4 \\
&\not \!\! F_{(5)} =  - f\,\mu \, (2 - 2\, {\cal N}_{1234} - 4\, (a^\dagger_3 
a^\dagger_4 a_3 a_4 + a^\dagger_1 a^\dagger_2 a_1 a_2))
\end{split}\end{equation}
where ${\cal N}_{1234}$ is the fermion number operator.
Since we are studying type IIB theory, 
both $\begin{pmatrix}\theta^1 \\ \theta^2\end{pmatrix}$ are in ${\bf 8_s}$. 
Using the foregoing, \eqref{eomthree} becomes:
\begin{equation}\label{eomfive}\begin{split}
\gamma_+ (\partial_{\tau} + \partial_\sigma)
\begin{pmatrix}\theta^1_1 \\ \theta^1_2 
\\ \theta^1_3 \\ \theta^1_4 \\ \theta^1_5 \\ \theta^1_6 \\ \theta^1_7
 \\ \theta^1_8
\end{pmatrix}
= - \half\, \gamma_+ \beta\mu\begin{pmatrix}
{\bar b}\, e^{-i\beta\mu\tau}(i\theta^1_7 - \theta^2_7) +{2f\over
\beta}\theta^2_1 
\\ {\bar b}\, e^{-i\beta\mu\tau}(i\theta^1_8 - \theta^2_8) +{2f\over
\beta}\theta^2_2 
\\ -{2f\over \beta}\theta^2_3
\\ -{2f\over \beta}\theta^2_4
\\ -{2f\over \beta}\theta^2_5
\\ -{2f\over \beta}\theta^2_6
\\ b\,e^{i\beta\mu\tau}(i\theta^1_1 + \theta^2_1)+ {2f\over \beta}\theta^2_7
\\ b\,e^{\beta\mu\tau}(i\theta^1_2 + \theta^2_2) + {2f\over \beta} \theta^2_8
\end{pmatrix} \,.
\end{split}\end{equation}
and 
\begin{equation}\label{eomsix}\begin{split}
\gamma_+ (\partial_{\tau} - \partial_\sigma)\begin{pmatrix}\theta^2_1 \\ 
\theta^2_2 \\ \theta^2_3 \\ \theta^2_4 \\ \theta^2_5 \\ \theta^2_6 \\
\theta^2_7 \\ \theta^2_8\end{pmatrix}
= \half\,\gamma_+ \beta\mu\begin{pmatrix}
 {\bar b}\, e^{-i\beta\mu\tau}(i\theta^2_7 + \theta^1_7) +{2f\over
\beta}\theta^1_1 
\\ {\bar b}\, e^{-i\beta\mu\tau}(i\theta^2_8 + \theta^1_8) +{2f\over
\beta}\theta^1_2 
\\ -{2f\over \beta}\theta^1_3
\\ -{2f\over \beta}\theta^1_4
\\ -{2f\over \beta}\theta^1_5
\\ -{2f\over \beta}\theta^1_6
\\ b\,e^{i\beta\mu\tau}(i\theta^2_1- \theta^1_1) +{2f\over \beta}\theta^1_7
\\ b\,e^{i\beta\mu\tau}(i\theta^2_2 - \theta^1_2) +{2f\over \beta}\theta^1_8
\end{pmatrix} \,,
\end{split}\end{equation}

One can solve this system of first order ordinary differential equations by 
Fourier expanding in the $\sigma$ coordinate and shifting each mode 
by a constant:
\begin{equation}\label{fourier}\begin{split}
\theta^1_\mu=exp( \coeff{1}{2}\,\varepsilon\, (i\beta\mu\tau +arg(b) ))\ 
\sum_{n=1}^\infty \alpha_{(n)}^{\mu} (\tau)\,  exp\Big ({in\sigma\over
\alpha' p^+ }\Big ) \\
\theta^2_\mu=exp( \coeff{1}{2}\,\varepsilon\,( i\beta\mu\tau +arg(b)))\ 
\sum_{n=1}^\infty \beta_{(n)}^{\mu} (\tau)\,  exp \Big({in\sigma\over
\alpha' p^+ }\Big ) \,, 
\end{split}\end{equation}
where $\varepsilon = -1 $ for $\mu=1,2$; $\varepsilon = +1 $ for $\mu=7,8$;
 and $\varepsilon = 0$ for all other  $\mu$.  
Shifting the phase by this and including the constant means
that in the equations we get $b,{\bar b}$  $\rightarrow |b|$ and 
all of the $e^{\pm i\beta\mu\tau}$ dependence cancels. 

The equations break into the three obvious groups. Eight of these fermions
are not affected by the B-field at all, while the others group as
($\theta^1_1,  
\theta^1_7, \theta^2_1, \theta^2_7$), 
($\theta^1_2, \theta^1_8, \theta^2_2, \theta^2_8$).    The mode
analysis is trivial for the fermions that do not couple
to the $B$-field, and we thus find eight sets of fermionic oscillators with 
frequencies:
\begin{equation}\label{fermfreqtwo}
\omega ~=~ \pm \, \sqrt{(f\, \mu)^2 ~+~ 
\Big({n \over \alpha' p^+}\Big)^2} \,,
\end{equation}
with each sign having multiplicity $4$. 

The groups ($\theta^1_1,  \theta^1_7, \theta^2_1, \theta^2_7$), 
($\theta^1_2, \theta^1_8, \theta^2_2, \theta^2_8$)
produce identical sets of equations so we need only consider one of them:
\begin{equation}\label{eomseven}\begin{split}
&\Big(\partial_\tau \mp \coeff{1}{2}\, i\beta\mu +  {i n \over \alpha' p^+}
\Big)\,  \begin{pmatrix}\alpha_{(n)}^{1} \\ \alpha_{(n)}^{7}\end{pmatrix} 
\\
& \hskip2cm =- \coeff{1}{ 2}\, |b|\beta \mu\, \begin{pmatrix}i\alpha_{(n)}^{7} -
 \beta_{(n)}^{7} \\ i\alpha_{(n)}^{1} + \beta_{(n)}^{1}\end{pmatrix}
- f \, \mu\, 
\begin{pmatrix}\beta_{(n)}^{1} \\ \beta_{(n)}^{7}\end{pmatrix} \,, 
\end{split}\end{equation}
\begin{equation}\label{eomeight}\begin{split} 
& \Big(\partial_\tau \mp\coeff{1}{2}\, i\beta\mu -
 {i n \over \alpha' p^+} \Big)\, 
\begin{pmatrix}\beta_{(n)}^{1} \\ \beta_{(n)}^{7}\end{pmatrix} 
\\
& \hskip2cm = \coeff{1}{ 2}\, |b|\, \beta \mu\,
\begin{pmatrix}i\beta_{(n)}^{7} + 
\alpha_n^7 \\ i\beta_{(n)}^{1} - \alpha_{(n)}^{1}\end{pmatrix}
+f \,\mu\, 
\begin{pmatrix}\alpha_{(n)}^{1} \\ \alpha_{(n)}^{7}\end{pmatrix} \,, 
\end{split}\end{equation}

Doing normal mode analysis we get,
\begin{equation}\label{normalmodes}\begin{split}
& \begin{pmatrix} - ip + {i\beta\mu \over 2} &- {  i|b|\beta \mu\over 2}  
&- f &{|b|\beta \mu\over 2} \\
- {i|b|\beta \mu\over 2} & - ip  - {i\beta\mu\over 2} &
- {|b|\beta \mu\over 2} &- f \\
f & {|b|\beta \mu\over 2} &ip +{i\beta\mu\over 2} &  
{i|b|\beta \mu\over 2} \\
-{|b|\beta \mu\over 2} & f &  {i|b|\beta \mu\over 2} &  ip -
 {i\beta \mu\over 2} \end{pmatrix}
\begin{pmatrix}
\alpha_{(n)}^{1} \\
\alpha_{(n)}^{7} \\
\beta_{(n)}^{1} \\
\beta_{(n)}^{7}
\end{pmatrix} \\
& \hskip9cm =
i\, \omega
\begin{pmatrix}
\alpha_{(n)}^{1} \\
\alpha_{(n)}^{7} \\
\beta_{(n)}^{1} \\
\beta_{(n)}^{7}\end{pmatrix}  ,  \end{split}\end{equation}
where $p \equiv  { n \over \alpha' p^+} $.
The four normal modes of this system therefore have frequencies:
\begin{equation}\label{eigenvalues}\begin{split}
&\pm \left[\bigg(\coeff{1}{2}\,  |b|^2\,\beta^2 \mu^2+f^2\mu^2+
p^2 + \coeff{1}{4}\, \beta^2 \mu^2 \bigg) \right. \\
& \hskip2cm \left.  \pm \coeff{1}{2}\,  {\beta \mu}\,
\sqrt{( 2f- |b|^2\beta)^2 \mu^2 + 4\,p^2(1+|b|^2)}\right]^{1/2}
\end{split}\end{equation}
If one uses the   equations of motion for the supergravity
background and the condition for enhanced supersymmetry: 
\begin{equation}\label{relationone} 
f^2 + \coeff{1}{4}\,\beta^2\, |b|^2 ~=~ 1\,, \qquad \beta ~=~ -2f \,,
\end{equation}
then the  frequencies simplify significantly:
\begin{equation}\label{fermfreqs}\begin{split}
\omega~=~ \pm\, \mu \pm \sqrt{\Big({n \over \alpha' p^+}\Big)^2 + \mu^2} \,,  
\end{split}\end{equation}
with all four permutations of the $\pm$ signs.

\end{appendix}
\providecommand{\href}[2]{#2}\begingroup\raggedright\endgroup

\begin{thebibliography}{10}

\bibitem{BerensteinJQ}
D.~Berenstein, J.~M.~Maldacena and H.~Nastase,
``Strings in flat space and pp waves from N = 4 super Yang Mills,''
JHEP {\bf 0204}, 013 (2002)
[arXiv:hep-th/0202021].
\bibitem{MetsaevBJ}R.~R.~Metsaev,``Type IIB Green-Schwarz
superstring in plane wave Ramond-Ramond  background,''Nucl.\ Phys.\
B {\bf 625}, 70 (2002)[arXiv:hep-th/0112044].
\bibitem{BlauNE}M.~Blau, J.~Figueroa-O'Farrill, C.~Hull and
G.~Papadopoulos,``A new maximally supersymmetric background of IIB
superstring theory,''JHEP {\bf 0201}, 047
(2002)[arXiv:hep-th/0110242].
%
\bibitem{BlauMW}
M.~Blau, J.~Figueroa-O'Farrill and G.~Papadopoulos,
``Penrose limits, supergravity and brane dynamics,''
Class.\ Quant.\ Grav.\  {\bf 19}, 4753 (2002)
[arXiv:hep-th/0202111].
\bibitem{GauntlettCS}
J.~P.~Gauntlett and C.~M.~Hull,
``pp-waves in 11-dimensions with extra supersymmetry,''
JHEP {\bf 0206}, 013 (2002)
[arXiv:hep-th/0203255].
\bibitem{MetsaevRE}
R.~R.~Metsaev and A.~A.~Tseytlin,
``Exactly solvable model of superstring in plane wave Ramond-Ramond  background,''
Phys.\ Rev.\ D {\bf 65}, 126004 (2002)
[arXiv:hep-th/0202109].
\bibitem{RussoRQ}J.~G.~Russo and A.~A.~Tseytlin,``On solvable
models of type IIB superstring in NS-NS and R-R plane wave
backgrounds,''JHEP {\bf 0204}, 021 (2002)[arXiv:hep-th/0202179].
\bibitem{SchwarzQR}J.~H.~Schwarz,``Covariant Field Equations Of
Chiral N=2 D = 10 Supergravity,''Nucl.\ Phys.\ B {\bf 226}, 269
(1983).
\bibitem{CorradoWX}
R.~Corrado, M.~Gunaydin, N.~P.~Warner and M.~Zagermann,
``Orbifolds and flows from gauged supergravity,''
Phys.\ Rev.\ D {\bf 65}, 125024 (2002)
[arXiv:hep-th/0203057].
\bibitem{KhavaevYG}A.~Khavaev and N.~P.~Warner,``An N = 1 supersymmetric Coulomb flow in IIB supergravity,''Phys.\ Lett.\ B {\bf 522}, 181 (2001)[arXiv:hep-th/0106032].
%
\bibitem{CPW}R.~Corrado, K.~Pilch and N.~P.~Warner,Work in progress.
%
\bibitem{PilchFU}K.~Pilch and N.~P.~Warner,``N = 1 supersymmetric renormalization group flows from IIB supergravity,''Adv.\ Theor.\ Math.\ Phys.\  {\bf 4}, 627 (2002)[arXiv:hep-th/0006066].
\bibitem{PilchEJ}K.~Pilch and N.~P.~Warner,``A new supersymmetric compactification of chiral IIB supergravity,''Phys.\ Lett.\ B {\bf 487}, 22 (2000)[arXiv:hep-th/0002192].
%
\bibitem{FreedmanGP}D.~Z.~Freedman, S.~S.~Gubser, K.~Pilch and N.~P.~Warner,``Renormalization group flows from holography supersymmetry and a  c-theorem,''Adv.\ Theor.\ Math.\ Phys.\  {\bf 3}, 363 (1999)[arXiv:hep-th/9904017].
\bibitem{KhavaevFB}A.~Khavaev, K.~Pilch and N.~P.~Warner,``New vacua of gauged N = 8 supergravity in five dimensions,''Phys.\ Lett.\ B {\bf 487}, 14 (2000)[arXiv:hep-th/9812035].
%
\bibitem{vanNieuwenhuizenWU}P.~van Nieuwenhuizen and N.~P.~Warner,``Integrability Conditions For Killing Spinors,''Commun.\ Math.\ Phys.\  {\bf 93}, 277 (1984).
%
\bibitem{LeighEP}R.~G.~Leigh and M.~J.~Strassler,``Exactly marginal operators and duality in four-dimensional 
N=1 supersymmetric gauge theory,''Nucl.\ Phys.\ B {\bf 447}, 95 (1995)[arXiv:hep-th/9503121].

\bibitem{GomisKM}
J.~Gomis and H.~Ooguri,
``Penrose limit of N = 1 gauge theories,''
Nucl.\ Phys.\ B {\bf 635}, 106 (2002)
[arXiv:hep-th/0202157].
\bibitem{ZayasRX}L.~A.~Zayas and J.~Sonnenschein,``On Penrose limits and gauge theories,''JHEP {\bf 0205}, 010 (2002)[arXiv:hep-th/0202186].
%
\bibitem{ItzhakiKH}N.~Itzhaki, I.~R.~Klebanov and S.~Mukhi,``PP wave limit and enhanced supersymmetry in gauge theories,''JHEP {\bf 0203}, 048 (2002)[arXiv:hep-th/0202153].
\bibitem{RomansAN}L.~J.~Romans,``New Compactifications Of Chiral
N=2 D = 10 Supergravity,''Phys.\ Lett.\ B {\bf 153}, 392
(1985).
\bibitem{KlebanovHH}I.~R.~Klebanov and E.~Witten,``Superconformal field theory on threebranes at a Calabi-Yau  singularity,''Nucl.\ Phys.\ B {\bf 536}, 199 (1998)[arXiv:hep-th/9807080].

\bibitem{CveticSI}
M.~Cvetic, H.~Lu and C.~N.~Pope,
``M-theory pp-waves, Penrose limits and supernumerary supersymmetries,''
Nucl.\ Phys.\ B {\bf 644}, 65 (2002)
[arXiv:hep-th/0203229].

\bibitem{CveticZS}M.~Cvetic, H.~Lu, C.~N.~Pope and K.~S.~Stelle,``T-duality in the Green-Schwarz formalism, and the massless/massive
 IIA  duality map,''Nucl.\ Phys.\ B {\bf 573}, 149
(2000)[arXiv:hep-th/9907202].
\bibitem{HassanBV}S.~F.~Hassan,``T-duality, space-time spinors and R-R fields in curved backgrounds,''Nucl.\ Phys.\ B {\bf 568}, 145 (2000)[arXiv:hep-th/9907152].
\bibitem{GimonSF}
E.~G.~Gimon, L.~A.~Pando Zayas and J.~Sonnenschein,
``Penrose limits and RG flows,''
arXiv:hep-th/0206033.

%
\bibitem{BrecherAR}
D.~Brecher, C.~V.~Johnson, K.~J.~Lovis and R.~C.~Myers,
``Penrose limits, deformed pp-waves and the string duals of N = 1 large N  gauge theory,''
JHEP {\bf 0210}, 008 (2002)
[arXiv:hep-th/0206045].

%
\end{thebibliography}
\end{document}